\documentclass{article}
\usepackage{spconf,amsmath,graphicx,url}

\title{INDEPENDENT VECTOR ANALYSIS WITH DEEP NEURAL NETWORK SOURCE PRIORS}
\name{Xi-Lin Li}
\address{GMEMS Technologies, Inc., 366 Fairview Way, Milpitas, CA 95035 (e-mail: lixilinx@gmail.com)}
\begin{document}
%
\maketitle
\begin{abstract}
This paper studies the density priors for independent vector analysis (IVA) with convolutive speech mixture separation as the exemplary application. Most existing source priors for IVA are too simplified to capture the fine structures of speeches. Here, we first time show that it is possible to efficiently estimate the derivative of speech density with universal approximators like deep neural networks (DNN) by optimizing certain proxy separation related performance indices. Experimental results suggest that the resultant neural network density priors consistently outperform previous ones in convergence speed for online implementation and signal-to-interference ratio (SIR) for batch implementation.
\end{abstract}

\section{Introduction}

Speech separation, also known as the cocktail problem, is a fundamental signal processing task. Although there is a surge of supervised neural network based speech separation studies recently, the unsupervised approaches, e.g., independent component analysis (ICA) based on the Infomax principle \cite{multi_sepa1} and independent vector analysis (IVA) \cite{iva_lap}, are still attractive due to their simplicity and low complexity, and the wide availability of multichannel recordings on today's end devices like smart phones, tablets, personal computers, smart speakers, and many more internet of things (IoT) devices. Probability density function (pdf) of speech is the key component driven the separation of mixtures in these unsupervised frameworks. The most widely adopted pdf models for speech are the multivariate Laplace and generalized Gaussian distributions \cite{pdf1, pdf2, pdf3}, either in the time or frequency domain. Specifically, previously studied multivariate source priors for IVA include the Laplace prior \cite{iva_lap}, non-spherical priors represented by chain-like overlapped cliques \cite{iva_prior_non_spherical, iva_prior_non_spherical1}, Student's t-distribution \cite{iva_prior_student}, generalized Gaussian distribution (GGD) \cite{iva_prior_ggd}, complex Gaussian scale mixture (CGSM) \cite{iva_prior_gsm} and Gaussian mixture model (GMM) priors \cite{iva_prior_gmm}. However, most of these source priors are too simplified to capture the fine structures of speeches. A finite mixture model (FMM) is expressive enough, but could be too complicated due to the need to estimate those nuisance parameters of FMM. Actually, only the separation of mixtures of two sources is considered with the GMM source prior in \cite{iva_prior_gmm}. Density estimation for multivariate random variable is known to be a hard problem due to the curse of dimensionality. Fortunately, as in most maximum likelihood (ML) estimation problems, ICA or IVA for speech separation only requires the derivative of density, which could be estimated with less difficulty in practice, as shown in this paper. Here, we choose the IVA framework for speech separation as it is implemented in the frequency domain, and is computationally cheaper than convolutive ICA implemented in the time domain. In the training phase, neural networks are used to approximate the derivative of speech density by optimizing certain proxy separation related objectives. In the test phase, these neural network source priors are fixed, and only the separation matrices are adapted. In this way, our source priors are expressive enough, and yet, learning rules for updating the separation matrices are kept to be simple. 

Before ending the introduction, we would like to briefly summarize the main contributions of our work and its { relation to prior work}. Our approach is not a supervised speech separation method, although both use the neural networks as universal approximators. Supervised speech separation attracts a lot of attentions recently. It typically assumes that the train and test mixtures are generated in similar fashions. Hence, the resultant black-box models can only be applied to very specific scenarios, e.g., the single and multiple channel separation methods in \cite{supervised_sepa2} and  \cite{supervised_sepa6}, respectively. On the other hand, IVA is a well formulated optimization problem. The same source prior can be useful in different mixing scenarios. One main contribution of this paper is to develop a practical approach for estimating the source priors in IVA with universal approximators like neural networks. Another main contribution is to experimentally demonstrate the performance gain of the resultant neural network priors over previous ones in a wide range of speech separation tasks.  

\section{Background}

\subsection{Mixing and Separation Models}

We assume that there are $N\ge 2$ speech sources and microphones. Recording of the $m$th microphone is expressed as
$
x_m(i) = \sum_{n=1}^N \sum_{j=0}^L  a_{mn}(j) s_n(i-j)$, where $1\le m\le N$, $i$ and $j$ are two discrete time indices, $a_{mn}(i)$ the room impulse response (RIR) from the $n$th source to the $m$th receiver, $L+1$ the length of RIR,  and $s_n(i)$ the $n$th source signal. It is convenient to rewrite the mixtures compactly as
$
\pmb x(i) = \sum_{j=0}^L \pmb A(j) \pmb s(i-j)
$, where $\pmb x(i) = [x_1(i), \ldots, x_N(i)]^T$, $\pmb s(i) = [s_1(i), \ldots, s_N(i)]^T$, $\pmb A(j)$ the mixing matrix, and superscript $T$ denotes transpose. Reversing the convolutive mixtures in the time domain can be computationally expensive. Hence, it is more popular to consider the mixing and separation models in the frequency domain as
$
\pmb X(\omega_k, t)  = \pmb H(\omega_k) \pmb S(\omega_k, t) $ and 
$
\pmb Y(\omega_k, t)  = \pmb W(\omega_k) \pmb X(\omega_k, t)$, where $1\le k\le K$, $K$ is the number of frequency bins, $\omega_k$ the discrete angular frequency index, $t$ the frame index, $\pmb H(\omega_k)$ the mixing matrix, $\pmb W(\omega_k)$ the separation matrices, $
\pmb S(\omega_k, t) =[S_1(\omega_k, t), \ldots, S_N(\omega_k, t)]^T 
$,
$
\pmb X(\omega_k, t)=[X_1(\omega_k, t), \ldots, X_N(\omega_k, t)]^T $, 
and 
$\pmb Y(\omega_k, t)=[Y_1(\omega_k, t), \ldots, Y_N(\omega_k, t)]^T $. Clearly, the frequency resolution need to be high enough in order to well approximate the linear convolution in the time domain as $K$ frequency domain instantaneous mixing processes. 

\subsection{IVA for ML Separation Matrix Estimation}

Let 
$ \pmb S_n(t)  = [S_n (\omega_1, t), S_n(\omega_2, t) , \ldots, S_n(\omega_K, t) ]^T  $
and
$
\pmb Y_n(t)  = [Y_n (\omega_1, t), Y_n(\omega_2, t) , \ldots, Y_n(\omega_K, t) ]^T 
$, where $1\le n\le N$.  
Note that $\pmb S_m(t)$ and $\pmb S_n(t)$ are two independent complex valued source vectors for $1\le m\ne n\le N$, hence the name IVA. IVA further assumes that $\pmb S_n(t_1)$ and $\pmb S_n(t_2)$ are independent for $t_1\ne t_2$, although  this might not be true in reality. Then, we can write the pdf of the observed  mixtures as
\begin{equation}
p_X[\pmb X(\omega_1), \ldots, \pmb X(\omega_K)] = \frac{\prod_{n=1}^N p_S(\pmb S_n)}{\prod_{k=1}^K |\det[\pmb H(\omega_k)]|^2}
\end{equation} 
where $|\det(\cdot)|$ denotes the absolute determinant of a square matrix,  $p_S(\cdot)$ the pdf of speech signal in the frequency domain, and we have omitted the frame index $t$ to simplify our writing. Hence, ML estimation for the separation matrices are given by the minimum of the following expected negative logarithm likelihood (NLL) function
\begin{align}\nonumber
& J(\pmb W(\omega_1), \ldots, \pmb W(\omega_K)) \\ \nonumber 
& \quad = E[-\log p_X[\pmb X(\omega_1), \ldots, \pmb X(\omega_K)] | \pmb W(\omega_1), \ldots, \pmb W(\omega_K) ] \\ \label{nll}
& \quad  = E[ -\sum_{n=1}^N \log p_S(\pmb Y_n) - \sum_{k=1}^K \log|\det[\pmb W(\omega_k)]|^2 ]
\end{align}
Thus, IVA turns to a well defined optimization problem given the form of source prior, i.e., $p_S(\cdot)$. Natural or relative gradient descent \cite{Amari96, Cardoso96} is the most popular optimization method for minimizing the NLL in (\ref{nll}). For batch processing, relative Newton method \cite{iva_prior_gsm, iva_opt_method1} and the auxiliary function technique for spherical source priors \cite{iva_opt_method2} are shown to converge fast. Here, we choose natural gradient descent as the optimizer since it is suitable for both online and batch implementations. The learning rate for separation matrices updating is bin-wisely normalized as the method proposed in \cite{norm_lr}. Hence, the only left piece to be solved is the source priors.     

\section{Deep Neural Network Priors for IVA}

\subsection{Neural Network Density Model for Speech}

Let us suppress indices $n$ and $t$, and simply write the density of $\pmb S = [S(\omega_1), \ldots, S(\omega_K)]^T$ as $p(\pmb S) = p[S(\omega_1), \ldots, S(\omega_K)]$. It is reasonable to impose two regularities on the possible forms of $p(\pmb S)$. First, $\pmb S$ must be circular in the sense that $p(\pmb S)$ only depends on the amplitudes of $S(\omega_k)$ for $1\le k\le K$, but not their phases. Second, $\pmb S$ must be sparse, i.e., ${\partial p(\lambda \pmb S)}/{\partial \lambda} \le 0$ for any $\pmb S$ and $\lambda > 0$. Then, $p(\pmb S)$ can only have form
\begin{equation}
-\log p(\pmb S|\pmb \theta) =  F(|S(\omega_1)|^2, \ldots, |S(\omega_K)|^2, \pmb\theta)
\end{equation}
where $\pmb \theta$ is a pdf parameter vector, and $F(\cdot)$ is a properly chosen function. Indeed, any such $F(\cdot)$ can define a valid pdf as long as $\exp(-F)$ is integrable. The sparsity regularity requires that 
\begin{equation}\label{der1}
\frac{\partial F(|S(\omega_1)|^2, \ldots, |S(\omega_K)|^2, \pmb\theta) }{\partial |S(\omega_k)|^2 } \ge 0, \quad 1\le k\le K
\end{equation}
Notice that minimizing the NLL in (\ref{nll}) only requires the following derivative,
\begin{equation}
-\frac{\partial \log p(\pmb S|\pmb \theta)}{\partial S^*(\omega_k)} = \frac{\partial F(|S(\omega_1)|^2, \ldots, |S(\omega_K)|^2, \pmb\theta) }{\partial |S(\omega_k)|^2 } S(\omega_k)
\end{equation}
where superscript $*$ denotes conjugate. Thus, all we need are the $K$ derivatives in (\ref{der1}), which could be approximated using a feedforward neural network (FNN) with nonnegative outputs.  

It is also possible to consider the temporal dependence among successive frames from the same source signal. Specifically, for Markov sources, we have 
\begin{equation}\label{p_h_t} 
p(\pmb S(t)|\pmb S(t-1), \ldots, \pmb S(1), \pmb \theta) = p(\pmb S(t)|\pmb h(t-1), \pmb \theta)
\end{equation}
where $\pmb h(t)$ is a hidden state vector at time $t$. We could use a recurrent neural network (RNN) with $K$ nonnegative outputs to model such densities as well. 

\subsection{Examples of Neural Network Density Priors} 

A neural network usually performs the best for normalized inputs. Here, we define the normalized spectrum vector as $\bar{\pmb S} =\pmb S/\|\pmb S\|$, where $\|\pmb S\|$ is the length of $\pmb S$. Amplitudes of its elements could be further compressed with an element-wise logarithm operation. We have tested the following neural network density model in our experiments
\[ -\frac{\partial \log p(\pmb S| \pmb h, \pmb \theta) }{\partial \pmb S^*}  = \log[1 + \exp ({\pmb \gamma})] \odot \bar{\pmb S} \]
with $\pmb \gamma$ as the output of the following three layered network
\begin{align}\nonumber
\pmb \alpha(t) & = \tanh( \pmb\Theta_1 [ \log |\bar{\pmb S}(t)|^2; \log\|\pmb S(t)\|; \pmb h(t-1); 1 ] ) \\ \nonumber
\pmb \beta(t) & = \tanh(\pmb\Theta_2 [ \pmb \alpha(t); 1 ] ) \\ \label{nn}
\pmb \gamma(t) & = \pmb\Theta_3 [ \pmb \beta(t) ; 1 ] 
\end{align}
where $\{\pmb\Theta_1, \pmb\Theta_2, \pmb\Theta_3\}$ are the model parameters, $|\cdot|$ takes the element-wise absolute value, $[\cdot\; ;\; \cdot]$ denotes stacking column vectors vertically, and hidden state vector $ \pmb h(t-1)$ is a subset of  $\pmb \alpha(t-1) $. Specifically, (\ref{nn}) defines a FNN when $\pmb h(t) = [ \,]$, and a RNN otherwise. The RNN model can only be used to update the separation matrices sequentially by keeping the temporal order, while the FNN one has no such limitation. It is possible to consider more complicated priors. Nevertheless, these simple ones perform competitively in our experiments.    

\subsection{Proxy Objective for Source Prior Estimation}

The separation results are determined by the source priors given the learning rules for separation matrices updating. Thus, it is possible to choose a proxy performance index measuring the goodness of separation, and `learn' the source priors to optimize the chosen proxy objective. In our experiments, we choose the following average permutation invariant (PI) absolute coherence as this objective
\begin{equation}\label{obj}
c(\pmb \theta) = \max_{\pi } \frac{1}{NK}\sum_{n=1}^N \sum_{k=1}^K \frac{ | E[ Y_{\pi(n)}(\omega_k, t) S^*_n(\omega_k, t) ] | }{\sqrt{ E[ |Y_{\pi(n)}(\omega_k, t)|^2] E[ |S_n(\omega_k, t)|^2 ] }}
\end{equation}  
where $\pi$ denotes an element of the set of all possible permutations of list $[1, 2, \ldots, N]$, $\pi(n)$ the $n$th element of permutation $\pi$, and we deliberately write $c(\pmb \theta)$ as a function of $\pmb\theta$ to show its dependence on the source prior parameter vector $\pmb\theta$. Similar PI objectives are used in supervised speech separation as well. Clearly, $c(\pmb \theta)$ is invariant to the scaling of separated outputs as well. In the training phase, the source signals are known. Thus, given the form of a source prior, we can optimize its parameters by maximizing the objective in (\ref{obj}) with deep learning tools like Pytorch. Such resultant estimated source prior implicitly defines a pdf suitable for the separation of speech mixtures. Note that unlike a FMM, there is no need to update the neural network priors in the test phase.   

\begin{figure}[htb]
	\begin{minipage}{.48\linewidth}
		\centering
		\centerline{\includegraphics[width=4cm]{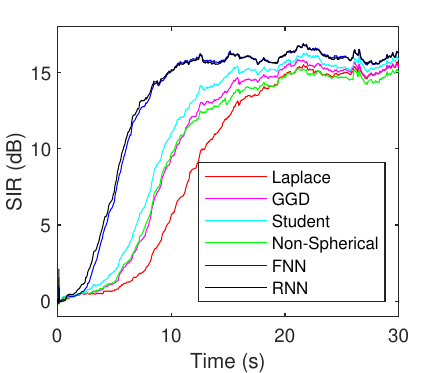}}
		\centerline{(a) Experiment 1}\medskip
	\end{minipage}
	\hfill
	\begin{minipage}{0.48\linewidth}
		\centering
		\centerline{\includegraphics[width=4cm]{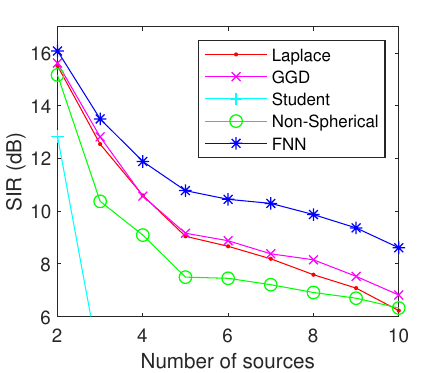}}
		\centerline{(b) Experiment 2}\medskip
	\end{minipage}
	\begin{minipage}{.48\linewidth}
		\centering
		\centerline{\includegraphics[width=4cm]{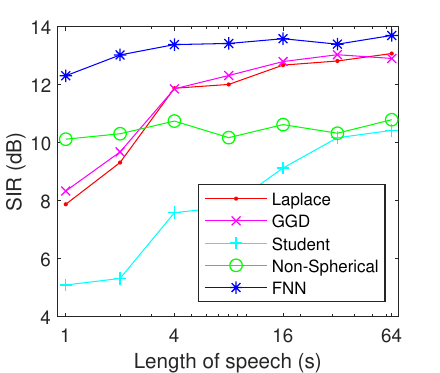}}
		\centerline{(c) Experiment 3}\medskip
	\end{minipage}
	\hfill
	\begin{minipage}{0.48\linewidth}
		\centering
		\centerline{\includegraphics[width=4cm]{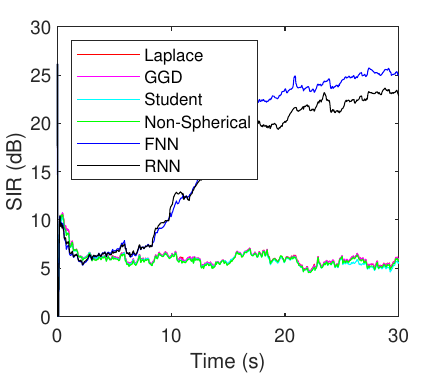}}
		\centerline{(d) Experiment 4}\medskip
	\end{minipage}
	\caption{Comparisons of different source priors in four separation tasks. Results are averaged over $50$ independent runs.}
	\label{fig:res}
\end{figure}

\section{Experimental Results}

Computer program reproducing the results reported below and sample separation results for subjective comparisons are available from our website \footnote{\url{https://github.com/lixilinx/IVA4Cocktail}}.

\subsection{General Setups} 

The training speeches are from a corpus of $100$ hour read LibriVox English books \cite{Librispeech}, and the test ones are from the well known TIMIT corpus. All have the same sampling rate, $16,000$ Hz. A short time Fourier transform (STFT) with frame size $512$ and hop size $160$ is used to convert the time domain signals to the frequency domain with analysis and synthesis windows designed by the method from \cite{fb}. This frequency resolution works well for separation of mixtures with low to moderate reverberations. All the separation matrices are initialized to the identity matrix.   

\subsection{The Training Environments} 

We have prepared one FNN and one RNN source prior. Dimensions of $\pmb\alpha$ and $\pmb\beta$ in (\ref{nn}) are the same, $512$. For the RNN model, the first $128$ elements of $\pmb \alpha$ serve as the hidden states. We always set $N=4$. Four randomly selected sources are artificially mixed as $ \pmb x(i) = \sum_{j=-16}^{16} {\pmb A(j)} \pmb s(i-j) /(1+|j|) $, 
where all the elements in $\pmb A(i)$ are standard Gaussian random variables. The normalized learning rate in natural gradient descent is set to $0.01$. Absolute coherence in the proxy objective of (\ref{obj}) is estimated over $128$ frames. We choose to reset the mixing matrices with a probability of $0.02$ after each evaluation of proxy objective. The simulation batch size is set to $64$. The preconditioned stochastic gradient method in \cite{psgd}  is used to optimize the neural network coefficients with default step size $0.01$ and a total of $20,000$ iterations. The final converged average absolute coherence is about $0.8$. 

\subsection{The Test Environments} 

The test speeches are convolutively  mixed through randomly generated RIRs using the image source method \cite{rir}. Sizes of the simulated room are $({\rm Length}=5,\, {\rm Width}=4,\, {\rm Height}=3)$, all in meters. Locations of simulated microphones are randomly and uniformly distributed inside of a sphere with radius $0.1$ and centered at $(2, 2, 1.5)$, while the positions of simulated speech sources are also equally distributed outside of a sphere with radius $1$ and the same center location. To simulate fractional delays, we first generate the RIRs with sampling rate $48,000$ and then decimate them to sampling rate $16,000$. The wall reflection coefficients are set to $0.25$ such that the typical converged signal-to-interference ratio (SIR) for the separation of two sources is about $15$ dB. This SIR number is also  representative for IVA tested on real world mixtures of two speeches recorded in living rooms with low to moderate reverberations.

\subsection{Test SIR Performance Comparisons}

We have designed four experiments to compare six source priors for speech separation, i.e., the Laplace one \cite{iva_lap}, GGD \cite{iva_prior_ggd}, Student's t-distribution \cite{iva_prior_student}, a non-spherical prior by grouping the bins into four cliques of equal size in Mel scale \cite{iva_prior_non_spherical}, and our estimated FNN and RNN source priors. The scaling ambiguity is resolved by the minimum distortion principle \cite{mdp}.     

{\em Experiment 1}: This experiment benchmarks the  convergence speed for online implementation. The separation matrices are updated once per frame with a fixed normalized learning rate. Here, we set $N=2$, and the normalized learning rate to $0.03$.  

{\em Experiment 2}: This experiment benchmarks the  statistical efficiency of different source priors in batch processing mode. Since the separation matrices are not necessarily updated sequentially in the temporal order, the RNN source prior is not considered. The recording length is $10$ s. We vary the number of sources. A total of $10,000$ separation matrix updatings are performed to ensure convergence before measuring the SIR performance. The normalized learning rate starts from $0.1$, and linearly reduces to $0.01$ at the end of iteration. 

{\em Experiment 3}: This is one more experiment comparing the statistical efficiency of different source priors in batch processing mode. Unlike Experiment 2, we set $N=3$, and vary the length of speeches. We also find that it is necessary to halve the initial normalized learning rate for the Student's t source prior to avoid occasional divergence. Other source priors do not suffer from such issue.  

{\em Experiment 4}: The last experiment compares the capacity of different source priors for correcting frequency permutations. Prior work and our experiences suggest that IVA might be trapped in local minima \cite{jbss}, and thus fails to solve the frequency permutation issue. One typical error pattern is to mix one source's high frequency band with another's low frequency band in a single separated output. Unfortunately, the SIR performance index is insensitive to such errors as most speech energy concentrates in low frequency band. To reliably reproduce this misbehavior, we consider a simple $2\times 2$ artificial mixing system consisting of low and high pass Butterworth filters as
$
\pmb A(z) = \begin{bmatrix}
(1+z^{-1})^2 & (1-z^{-1})^2\\
(1-z^{-1})^2 & (1+z^{-1})^2
\end{bmatrix} /(1+0.17z^{-2})
$. High frequency energy is emphasized by passing the outputs through filter $1-z^{-1}$ before measuring the SIR. Other settings are the same as that of Experiment 1.

Fig.~1 summarize the experimental results. Experiment 1 suggests that the neural network priors lead to the fasted convergence. The RNN model only delivers a marginal performance gain over the FNN one. The Student's t prior performs the best among those simple ones, confirming the observations in \cite{iva_prior_student}. Both Experiment 2 and 3 suggest that the FNN source prior is significantly more efficient than previous ones for speech separation when the number of sources is large or length of speech is short. Among those simple priors, Laplace and GGD show similar performance. Still, the GGD prior seems perform slightly better than the Laplace one. This observation is consistent with those in \cite{iva_prior_ggd}. The non-spherical source prior performs better than other simple ones only when the length of speech is small. Its performance might be sensitive to the definition of cliques  \cite{iva_prior_non_spherical, iva_prior_non_spherical1}, and our definition is not necessarily optimal for all these tasks. Performance of the Student's t prior can be improved with smaller learning rates and more iterations. But, it is still less competitive than other simple ones in Experiments 2 and 3. Lastly, Experiment 4 suggests that only the neural network source priors are able to solve the low and high frequency bands permutation issue. This is not astonishing since none of the other simple source priors can capture the fine structures of speeches.

\section{Conclusion}

Separation of speech mixtures is a longstanding challenging signal processing problem. Speech density model is the key component in unsupervised separation frameworks like the independent vector analysis (IVA). In this paper, we have shown that it is possible to efficiently estimate the derivative of density of speeches represented in the frequency domain by optimizing certain separation related proxy objectives like the absolute coherence between source signals and separated outputs. Specifically, we have considered neural network speech density priors with heuristic design constraints like circularity and sparsity. Experimental results confirm that these deep neural network source priors considerably outperform previous ones in convergence speed for online implementations and statistical efficiency in batch processing mode.

\end{document}